\documentclass[aps,prb,twocolumn,showpacs]{revtex4}
\usepackage{graphicx}
\usepackage{amssymb}

\begin{document}


\title{Complex interplay of 3$d$ and 4$f$ magnetism in La$_{1-x}$Gd$_x$MnO$_3$}

\author{J. Hemberger$^1$, S. Lobina$^1$, H.-A.~Krug von Nidda$^1$, N.~Tristan$^1$, V.Yu.~Ivanov$^2$, A.A.~Mukhin$^2$,
A.M.~Balbashov$^3$, and A. Loidl$^1$}
\address{%
$^1$Experimentalphysik V, Center for Electronic Correlations and Magnetism, \\
Institut f\"{u}r Physik, Universit\"{a}t Augsburg, D-86135 Augsburg, Germany \\ %
$^2$ General Physics Institute of the Russian Academy of Sciences,
38 Vavilov Street, 119991 Moscow, Russia \\ %
$^3$ Moscow Power Engineering Institute, 14 Krasnokasarmennaja Street, 111250 Moscow, Russia %
}

\begin{abstract}
We report on structural, magnetic, electrical, and thermodynamic
properties of Gd-doped LaMnO$_3$ single crystals for Gd doping
levels $0 \leq  x  \leq 1$. At room temperature, for all doping
levels the orthorhombic O' phase is indicative for a strong
Jahn-Teller distortion. All compositions are insulating. The
magnetism of La$_{1-x}$Gd$_x$MnO$_3$ is dominated by the
relatively strong Mn-O-Mn superexchange. For increasing Gd doping
The weakening of the nearest-neighbor exchange interactions due to
the significant decrease of the Mn-O-Mn bond angles leads to the
continuous suppression of the magnetic phase-transition
temperature into the A-type antiferromagnetic low-temperature
phase. The temperature dependence of the magnetization can only be
explained assuming canting of the manganese spins. The magnetic
moments of Gd are weakly antiferromagnetically coupled within the
sublattice and are antiferromagnetically coupled to the Mn
moments. For intermediate concentrations compensation points are
found, below which the spontaneous magnetization becomes negative.
In pure GdMnO$_3$ the Mn spins undergo a transition into a
complex, probably incommensurate magnetic structure at 41.5~K,
followed by a further ordering transition at 18-20~K revealing
weak ferromagnetism due to canting and finally by the onset of
magnetic order in the Gd sublattice at 6.5~K. At the lowest
temperatures and low external fields both magnetic sublattices
reveal a canted structure with antiparallel ferromagnetic
components.
\end{abstract}



\maketitle

\section{Introduction}

Exactly since one decade, stimulated by the observation of
colossal magnetoresistance effects by von Helmolt et al.\cite{1}
and Chahara et al.\cite{2} the physics of doped manganese
perovskites is in the focus of solid-state research. The enormous
progress in this field is documented in recent review articles by
Coey et al.,\cite{3} Nagaev,\cite{4} and Salomon and
Jaime.\cite{5} In addition to their technological potential the
manganites also represent an enormous and fascinating playground
for basic solid-state research: unconventional and exotic ground
states are established via the competing interplay of the internal
degrees of freedom, like spin, orbital momentum, charge, as well
as lattice degrees of freedom.\cite{13}

In this communication we report on structural details, on
susceptibility, magnetization and heat capacity of single
crystalline La$_{1-x}$Gd$_x$MnO$_3$ and, hence, our focus is
directed towards the isovalent doping of the A-site in $R$MnO$_3$,
with $R$ denoting the rare-earth elements lanthanum (4f$^0$) or
gadolinium (4f$^{7}$). From a general point of view, the
rare-earth manganites can be grouped into two classes: the
compounds from La to Dy (with the exception of Ce and Pm),
characterized by larger ionic radii, reveal an orthorhombic
perovskite-derived structure, while the compounds from Ho to Lu
crystallize in a hexagonal structure, not related to the
perovskites.\cite{6} Focusing on the first group, when going from
La to Dy manganite, the tolerance factor decreases with the
decrease of ionic radii and the buckling and tilting of the ideal
cubic structure becomes stronger.\cite{ima98} Consequently the
Mn-O-Mn bond angles are reduced resulting in a strong suppression
of the antiferromagnetic (AFM) ordering temperature of the Mn
subsystem, from 140 K for LaMnO$_{3}$ to 40 K for TbMnO$_{3}$ and
the corresponding transformation of the Mn magnetic structure from
an antiferromagnetic layer A-type ordering with a weak
ferromagnetism (A$_y$F$_z$) (La, Pr, Nd) \cite{6} to an
incommensurate sinusoidal structure (Tb).\cite{8}  At the same
time, the Jahn-Teller (JT) transition, i.e.\ the orbital-order
transition, which is well below 1000~K in LaMnO$_3$, is shifted to
values of almost 1500~K in TbMnO$_3$.\cite{7a} However, strong
hysteresis effects, domain reorientation, history dependence,
oxygen diffusion, and oxidation processes do not allow a precise
determination of the JT transition temperatures.\cite{7a}

The magnetic moments of the rare-earths ions are polarized due to
the coupling with  the Mn subsystem resulting in a noticeable
anisotropic contribution to the low-temperature magnetic and
thermodynamic properties of the manganites as we have recently
reported for PrMnO$_3$ and NdMnO$_3$.\cite{10}  The coupling
within the rare-earth sublattice is antiferromagnetic and rather
weak. In many cases the rare-earth moments do not exhibit
long-range magnetic order (and if, the ordering temperatures are
below 10~K). The magnetic structure in the rare-earth manganites
can be rather complex, as it has been recognized already long time
ago. In TbMnO$_3$ Quezel et al.\cite{8} found a sine-wave ordering
of the Mn$^{3+}$ moments with the ordering wave vector along the
$b$-axis below 40~K and a short-range incommensurate ordering of
the Tb$^{3+}$ moments with a different wave vector below 7~K.
Similar magnetic structures were also observed in orthorhombic
HoMnO$_{3}$. \cite{8a, munoz} Metamagnetic behavior has been
detected in the Gd, Tb and Dy compounds.\cite{9}


Turning now specifically to GdMnO$_3$, it has first been
synthesized by Bertaut and Forrat\cite{20} and by Belov et
al.\cite{21} The magnetic properties of pure GdMnO$_3$
were studied and have been reported previously by Troyanchuk et
al.\cite{9} It was speculated that GdMnO$_3$ undergoes a smeared
out phase transition into a complex magnetic order of the
manganese sublattice below 40~K with the Gd sublattice
antiferromagnetically ordered with respect to the Mn moments and
that antiferromagnetic order within the Gd sublattice takes place
below 6~K. In addition, a metamagnetic transition has been
observed in external fields of 5~kOe.\cite{9}
Based on susceptibility and M\"ossbauer results, Zukrowski et
al.\cite{24} concluded that the Mn moments order at 40~K, while
long-range magnetic order in the Gd sublattice is established
below 20~K. Very recently several contributions concerning pure
perovskitic rare earth manganites (containing pure LaMnO$_3$ and
GdMnO$_3$) were published studying the interrelation between
structural constraints and the magnetic and orbital structures.
Both, N\'eel and orbital (JT) transition systematically depend on
the in-plane Mn-O-Mn bond angle $\phi$. The findings are
interpreted in terms of spin frustration\cite{kimura03} and
orbital order-disorder transitions.\cite{zhou03}

In this work we provide a detailed study of the structural,
magnetic and electrical properties of La$_{1-x}$Gd$_x$MnO$_3$. We
use this system as a systematic variation of the Mn-O-Mn angle and
relate it to the development of the complex magnetic groundstates.
In addition, we provide heat-capacity results for GdMnO$_3$ as
function of temperature and field to unravel the complex sequence
of magnetic phase transitions in this compound and to derive a
detailed $(H,T)$-phase diagram.

\section{Experimental Details}

La$_{1-x}$Gd$_x$MnO$_3$ single crystals were grown in Ar flow by a
floating-zone method with radiation heating for Gd concentrations
$x = 0, 0.25, 0.5, 0.75$ and 1. X-ray powder-diffraction
experiments were performed on powder of crushed single crystals at
room temperature with a STOE diffractometer utilizing
Cu-K$_{\alpha}$ radiation with a wave length $\lambda =
0.1541$~nm. The magnetic susceptibility and the magnetization were
recorded using a commercial SQUID magnetometer for temperatures $T
<$~400~K and external magnetic fields up to 50~kOe. The electrical
resistance has been measured using standard four-probe techniques
in van-der-Pauw geometry\cite{pauw58} in home-built cryostats from
1.5~K to room temperature and in bar-type geometry in a home-built
oven up to 1200~K. The specific heat was recorded in a home-built
equipment using ac methods for temperatures $T < 200$~K and
external magnetic fields up to 140~kOe.

The samples with $x=1$ and 0.75 revealed a strong magnetic
anisotropy with the easy direction of magnetization along the
$c$-axis. The $x=0.5$ and 0.25 samples exhibited a weaker
anisotropy probably due to the twin structure, resulting in an
appearance of a spontaneous magnetization not only in the easy
direction but also in a hard one. The data shown below, except a
few ones for pure GdMnO$_3$, correspond to the measurements along
the easy direction.

\section{Results and discussion}

\begin{figure}[tb]
\includegraphics[clip,width=65mm]{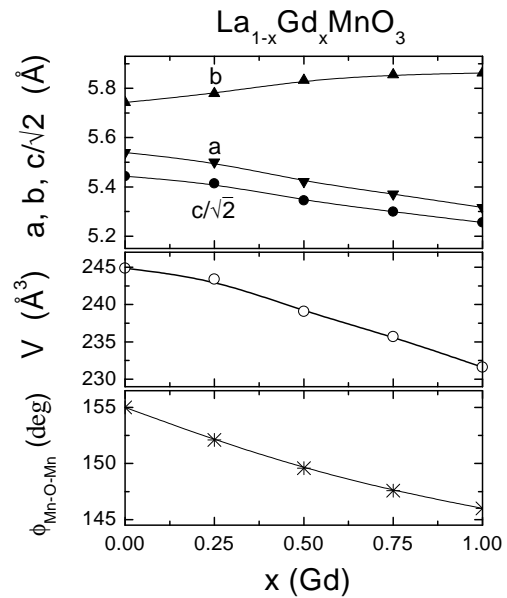}
\caption{Lattice constants $a$, $b$, and $c/\sqrt{2}$ (upper
frame), volume of the unit cell (middle frame), and in-plane
Mn-O-Mn bond angle $\phi$ in La$_{1-x}$Gd$_x$MnO$_3$ vs.\ Gd
concentration $x$.
\label{fig1}}
\end{figure}

From x-ray diffraction all samples investigated revealed the O'
orthorhombic structure (Pbnm, JT distorted, see i.e.\
\onlinecite{landolt} for the canonical convention of the
structural nomenclature). No impurity phases were detected above
background level. The diffraction patterns were refined using
Rietveld analysis. The lattice constants and the volume of the
unit cell as derived from the profile analysis are shown in
Fig.~\ref{fig1}. For all concentrations we find $b > a >
c/\sqrt{2}$ indicative for a static JT distortion superposed to
the high temperature O-type (i.e.\ not JT distorted) orthorhombic
structure which results from the buckling and tilting of the
MnO$_6$ octahedra due to geometrical constraints. On increasing Gd
concentration the lattice constants $a$ and $c$ are slightly
decreasing, while $b$ increases continuously. From this
observation some remarkable structural details can be deduced: The
parameter $\varepsilon = (b - a)/(a + b)$ characterizes the
orthorhombic distortion. This distortion is continuously
increasing on Gd doping due to the substitution of La ions by the
smaller Gd. In addition, the increasing inequality of the lattice
constants $a$ and $b$
with increasing $x$ signals an increasing tilting of the octahedra
around the $b$-axis and implies significant deviations of the
Mn-O-Mn bond angle $\phi$ within th $ab$-plane from $180^\circ$.
In the lowest frame of Fig.~\ref{fig1} we illustrate the
concentration dependence of the in-plane Mn-O-Mn bond angle $\phi$
as estimated from the Rietveld refinement of the x-ray powder
data.\cite{y-parameter}
The G-type orbital order of LaMnO$_3$ determines the magnetic
superexchange interactions between the Mn spins. The alteration of
the orbital configuration via the Mn-O-Mn bond angle should
influence in the magnetic properties of the spin system.
According to the Kanamori-Goodenough rules (assuming $180^\circ$
bond angles) the semi-covalent orientation of the $e_g^1$ orbitals
within the $ab$-plane leads to ferromagnetic (FM) superexchange
interactions, the covalent orientation of these orbitals along
$c$-direction leads to AFM superexchange.\cite{goo63} This results
into the well known A-type AFM order with FM $ab$-planes coupled
antiferromagnetically along the $c$-axis. The deviation of the
in-plane Mn-O-Mn bond angles from $180^\circ$ will weaken the FM
interactions within the $ab$-planes but not alter
the AFM coupling along $c$. The paramagnetic Curie-Weiss
temperature $\theta$ is determined by the sum of FM in-plane
interactions and AFM coupling along $c$. Concomitantly we expect a
decrease of the N\'eel temperature for the manganese sublattice,
determined by the sum of the absolute values of the different
interactions.

\begin{figure}[tb]
\includegraphics[clip,width=65mm]{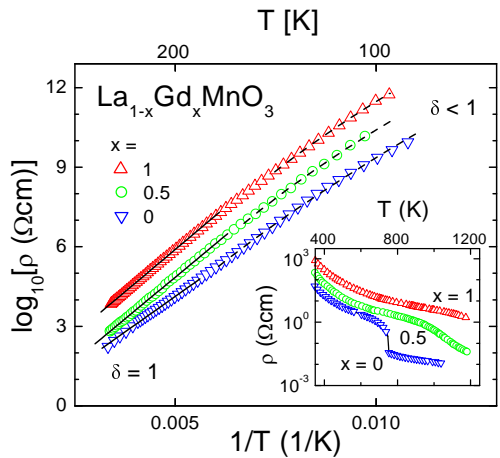}
\caption{Temperature dependence of the resistivity in
La$_{1-x}$Gd$_x$MnO$_3$ for various compounds plotted in an
Arrhenius representation. (The $x=1$ curve is shifted for better
clarity.) The solid lines at high temperatures represent Arrhenius
fits indicating a purely activated type of behavior of the charge
carriers. The dashed lines at low temperatures represent
variable-range-hopping conductivity. Inset: Resistivity above
400~K. %
\label{fig10}}
\end{figure}

\begin{figure}[b]
\includegraphics[clip,width=65mm]{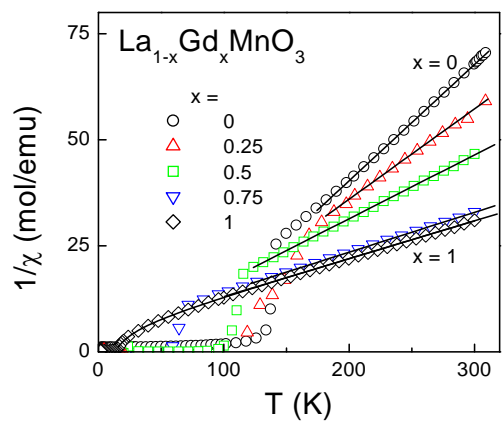}
\caption{Inverse dc susceptibility ($\chi = M/H$) versus
temperature as measured in an external magnetic field of 10~Oe.
The solid lines through the data for $x = 0, 0.25, 0.5$ and 0.75
represent fits to Curie-Weiss laws. The solid line through the
data for $x = 1$ indicates the result of a fit taking into account
two magnetic sublattices coupled ferrimagnetically.
\label{fig2}}
\end{figure}

The ratio $c/a< \sqrt{2}$ (see Fig.\ \ref{fig1}) almost remains
constant. This indicates the absence of qualitative differences
concerning the driving force of the JT distortion in all the
compounds investigated. The reason is that the substitution of
isovalent Gd$^{3+}$ for La$^{3+}$ gives Mn$^{3+}$ in an octahedral
environment for all Gd concentrations $x$, with one electron in
the upper e$_g$ doublet of the 3$d$ manifold, which is JT active.
As a consequence the transport properties do not change their
insulating characteristics for all Gd concentrations.
Fig.~\ref{fig10} shows the resistivity for the pure compounds and
for $x=0.5$ in an Arrhenius representation. All samples show
insulating behavior with a relatively high room-temperature
resistivity of $\approx 3$~k$\Omega$cm (again demonstrating the
good sample stoichiometry). At high temperatures the resistance of
all samples can be described by a simple thermally activated
behavior, indicated by straight solid lines in Fig.~\ref{fig10}.
The energy barriers are of the order of 0.2 to 0.25~eV, not
systematically depending on concentration. At low temperatures
significant deviations from an Arrhenius type of resistance
behavior can be observed. In a temperature range below 150~K
three-dimensional variable-range hopping following $\ln \rho
\propto (T_0/T)^{\delta}$, with values of $T_0$ of the order of
$5\times 10^9$~K for $\delta=1/4$, values typically observed in
manganites.\cite{26} It has to be stated that in the examined
temperature range accessible below 150~K (limited by the upper
value of the measurable impedance range) similar good fitting
results can be achieved using an exponent of $1/2$ indicating the
opening of a Coulomb gap and, therefore, no final conclusions on
the influence of correlation effects on the hopping mechanisms can
be drawn from this work.\cite{seeger} However, from the
resistivity results it seems that the electron-band structure is
not altered significantly upon Gd doping. Hence, the JT ordering
temperature is only influenced by the local lattice distortions of
the oxygen octahedron due to geometrical constraints.  And indeed,
for pure LaMnO$_3$ the Jahn-Teller transition has been reported to
occur at $T_{\rm JT} = 750$~K\cite{24,25}, while it is shifted to
950~K in pure PrMnO$_3$.\cite{14} The inset of Fig.~\ref{fig10}
shows resistivity data for the pure compounds LaMnO$_3$ and
GdMnO$_3$ and the mixed compound La$_{0.5}$Gd$_{0.5}$MnO$_3$ in a
temperature range from 400 to 1200~K. In LaMnO$_3$ (triangles
down) the JT transition can clearly be seen as a sharp jump in the
resistivity curve. In GdMnO$_3$ no such feature can be detected
demonstrating that the JT transition occurs even above $T \approx
1200$~K. The increase of the JT transition temperature with
decreasing Mn-O-Mn bond angle is in accordance with recent
studies.\cite{kimura03,zhou03}

\begin{figure}[b]
\includegraphics[clip,width=65mm]{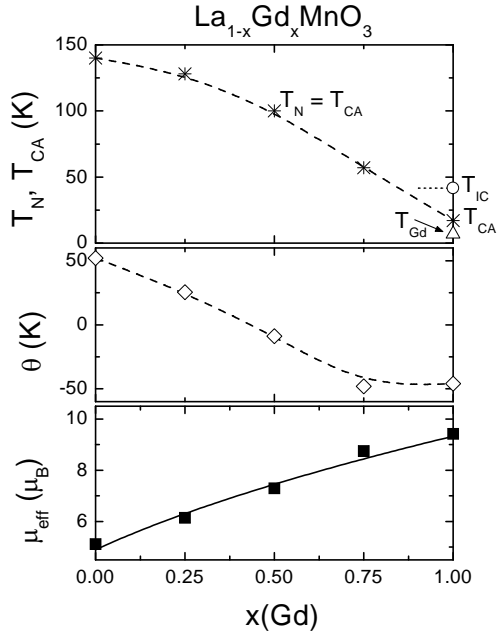}
\caption{Concentration dependencies of the parameters as obtained
from the analysis of the susceptibility measurements: N\'eel
temperatures (asterixes: upper frame), Curie-Weiss temperatures
(open rhombs: middle frame), and paramagnetic moments (solid
circles: lower frame). The solid line through the concentration
dependence is a fit as described in the text. The dashed lines are
drawn to guide the eye.
\label{fig3}}
\end{figure}

Fig.~\ref{fig2} shows the inverse magnetic susceptibilities for
all concentrations investigated. At elevated temperatures all
susceptibilities follow a Curie-Weiss (CW) law. The Curie constant
is increasing due to the increasing contribution of the rare-earth
moment (Gd$^{3+}$: spin-only, $S = 7/2$). As expected due to the
decreasing bond angles (see Fig.~\ref{fig1}) between neighboring
Mn ions (Mn$^{3+}$: high-spin, $S = 2$) the N\'eel temperatures
$T_{\rm N}$ are significantly decreasing. The dependence of
$T_{\rm N}$ on the Mn-O-Mn bond angle fits well with the findings
of Kimura et al.\ \cite{kimura03} for other $R$MnO$_3$ compounds.
As we will show in the following, with the exception of pure
GdMnO$_3$, no phase transition into a long-range AFM order of the
Gd moments can be detected in the temperature range investigated.
The concentration dependencies of the effective moments (lower
frame), the CW temperatures $\theta$ (middle frame), and the
magnetic transition temperatures (upper frame) are shown in
Fig.~\ref{fig3}. The onset of A-type AFM order at $T_{\rm CA}$ for
the doped compounds is well indicated via the observation of a
ferromagnetic component, due to spin canting (see
later).\cite{solovyev} As stated above, $T_{\rm CA}$ continuously
decreases from 140~K in LaMnO$_3$ to $\approx 20$~K in GdMnO$_3$.
But in the pure Gd compound additional magnetic transitions into
an incommensurate phase around $T_{IC}\approx 40$~K and into an
AFM order of the Gd moments can be detected by dc- and
ac-susceptibility measurements, as well as by heat-capacity
experiments, which will be discussed later. The positive CW
temperature $\theta$ of pure LaMnO$_3$, which is an A-type
antiferromagnet below 140~K, signals the importance of the
ferromagnetic in-plane super-exchange interactions. $\theta$
continuously decreases with increasing Gd concentration.
It is interesting to note that the CW temperatures become negative
for Gd concentrations $x \geq 0.5$. This signals the decreasing
importance of the ferromagnetic in-plane exchange on increasing
$x$. As already discussed above the concentration dependent
decrease of $T_{\rm CA}$ is related to the decreasing Mn-O-Mn
bond-angles altering the orbital overlap. Obviously this effect
influences particularly the FM superexchange interactions within
the $ab$-plane, as already suggested in literature.\cite{zhou03}
The decreasing importance of the dominating ferromagnetic in-plane
exchange enhances the influence of the AF next nearest-neighbor
exchange in the $ab$-plane and promotes a change of the spin
structure for the Mn sublattice, as will be discussed later for
pure GdMnO$_3$. However, the paramagnetic moments can well be
approximated assuming $\mu_{eff}^2 /(g\mu_B)^2 = x \times
S_{Gd}(S_{Gd} + 1) + S_{Mn}(S_{Mn} + 1)$ (solid line in the lower
frame of Fig.~\ref{fig3}).

\begin{figure}[tb]
\includegraphics[clip,width=65mm]{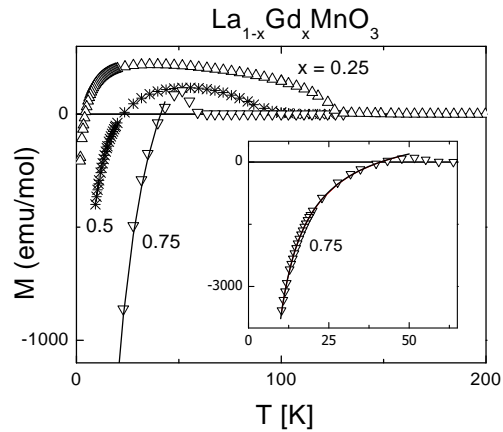}
\caption{Magnetization $M$ vs.\ temperature in
La$_{1-x}$Gd$_x$MnO$_3$ for $x = 0.25$ (triangles up), 0.5
(stars), and 0.75 (triangles down). The data were recorded on
cooling from room temperature in an external magnetic field of
10~Oe. The solid line represents a fit which takes into account
the paramagnetic contribution of the Gd subsystem polarized by the
internal field of the manganese moments due to Gd-Mn coupling.
Inset: Data and fit for $x = 0.75$ on an enlarged scale.
\label{fig4}}
\end{figure}

Fig.~\ref{fig4} provides a closer look on the magnetization of
La$_{1-x}$Gd$_x$MnO$_3$ for concentrations $x = 0.25, 0.5$, and
0.75 as measured in fields of 10~Oe.  At the magnetic ordering
temperature we observe a spontaneous ferromagnetic moment which
results from canting of the manganese spins. For the compounds
with Gd concentrations $x = 0.25, 0.5$ and 0.75 we find clear
indications for compensation points, the temperature of which
increases on increasing $x$. Via the antiferromagnetic coupling of
the Gd spins to the manganese moments, the Gd spins are polarized
antiparallel with respect to the FM component of the manganese
spins, which can be explained by Dzyaloshinsky-Moriya interaction
and is an intrinsic feature of the A-type AFM state in $R$MnO$_3$
systems.\cite{solovyev} With decreasing temperatures the Gd spins
are more and more aligned in the exchange field of the manganese
moments, following a Curie-type temperature dependence and finally
yielding a negative magnetization at the lowest temperatures, when
the polarization of the Gd spins exceeds the FM component of the
Mn moments. The solid line in the inset of Fig.~\ref{fig4} shows
that the negative magnetization $M$ can well be approximated by $M
= M_{Mn} + (H_{int} + H_{ext})\times C_{Gd}/T$ , where the
spontaneous magnetization $M_{Mn}$ and the internal field
$H_{int}$ result from the Mn moments and the external field is
$H_{ext} = 10$~Oe. (Note that $M_{Mn}$ was used as a fitting
parameter and considered to be constant well below $T_{\rm CA}$.)
$C_{Gd}$ is the Curie constant of the Gd moments. The best fit was
obtained using an internal magnetic field of the order of 8~kOe
created by the moments of the manganese sublattice.

\begin{figure}[b]
\includegraphics[clip,width=65mm]{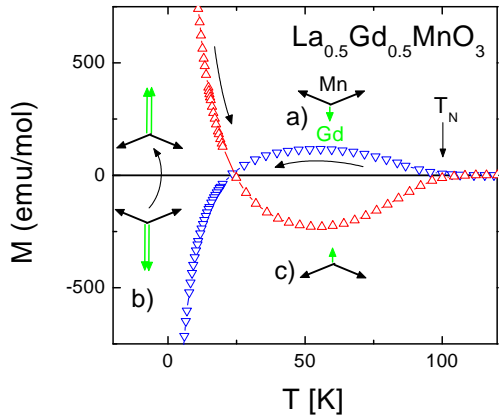}
\caption{Magnetization $M$ vs.\ temperature for
La$_{0.5}$Gd$_{0.5}$MnO$_3$ as measured in an external field of
10~Oe. The data were taken on cooling from room temperature
(triangles down). At the lowest temperatures the magnetization was
switched in an external field of 50~kOe. Then $M$ was measured on
heating (triangles up). The insets a), b), and c) indicate the
spin structure at the respective temperatures and cycle positions.
\label{fig5}}
\end{figure}

Fig.~\ref{fig5} shows the magnetization $M(T)$ for
La$_{0.5}$Gd$_{0.5}$MnO$_3$ in more detail for heating and
cooling. The temperature cycles are indicated by arrows. Triangles
down denote the cooling cycle in an external field of 10~Oe. At
$T_{\rm N} = 95$~K, the onset of a spontaneous magnetization along
$c$ indicates the appearance of a canted antiferromagnetic
structure of the Mn moments. Compensation due to the AFM alignment
of the Gd spins occurs close to 25~K. In the magnetization cycle,
the magnetization now was switched to positive values by an
external magnetic field of 50~kOe, the field was switched off
again  and then the sample was heated in a field of 10~Oe.  As
expected $M$ is exactly reversed. The spin structures in the
ordered phases including the spin reversal are also indicated in
Fig.\ \ref{fig5}.

\begin{figure}[tb]
\includegraphics[clip,width=65mm]{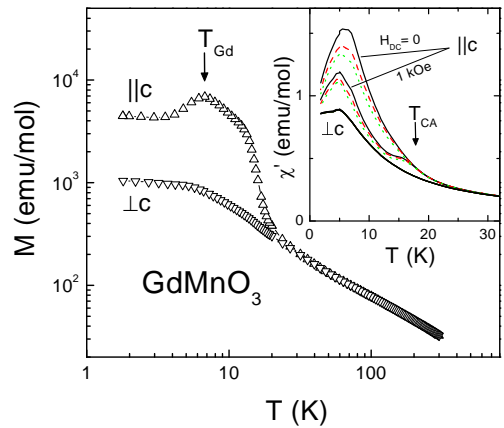}
\caption{Magnetization $M$ vs.\ temperature for GdMnO$_3$ plotted
on a double-logarithmic scale as measured in an external field of
1000~Oe. The easy (triangles up) and the hard direction (triangles
down) are shown which split approximately at 20~K. The inset shows
the ac susceptibility measured along $c$ and within the $ab$-plane
at a dc field of zero and 1000~Oe at 3 measuring frequencies of 1
(solid line), 35 (dashed line), and 1000~Hz (dotted line).
Perpendicular to the c-axis no dependence on magnetic bias field
or frequency can be detected.
\label{fig6}}
\end{figure}

Turning now to the magnetism of pure GdMnO$_3$, Fig.~\ref{fig6}
shows magnetization vs.\ temperature on a double-logarithmic
scale. Above 100~K the magnetization reveals the characteristics
of a pure paramagnet and follows a Curie-Weiss law with a CW
temperature $\theta = -35$~K. Down to 20~K the magnetization can
well be described using a fit formula for the high-temperature
susceptibility ($\chi=M/ H$) taking both sublattices into account
(see page 87 in Ref.\ \onlinecite{goo63}) as shown as solid line
in Fig.~\ref{fig2}. Finally at $T_{\rm CA}\approx20$~K the
magnetization reveals the onset of a significant anisotropy with a
strong and abrupt increase in the easy direction ($c$-axis) due to
the appearance of spontaneous magnetization in the canted state,
while $M$ only increases weaker and continuously along the hard
direction ($\bot c$). A smeared-out peak indicates the onset of
AFM long-range order of the Gd sublattice at $T_{\rm Gd}\approx
7$~K. Corresponding results can be found in the ac susceptibility
shown in the inset of Fig.~\ref{fig6}. The anisotropic splitting
of $\chi_{ac}$ parallel and perpendicular to the $c$-direction can
clearly be observed. Below $T_{\rm CA}$ a weak frequency
dependence for the susceptibility along $c$ appears. In addition a
shoulder at 18~K emerges in a magnetic bias field of 1~kOe. Both,
field and bias dependence cannot be found perpendicular to the
$c$-direction and can be explained by the dynamics of domains
carrying a FM component in $c$-direction, present in the canted
AFM state, and corresponding saturation effects. Towards lower
temperatures a peak or kink follows at $\approx 7$~K. Based on
these susceptibility results one is tempted to deduce an ordering
temperature of $\approx20$~K for the Mn and 7~K for the Gd
sublattice. But that the sequence of magnetic ordering transitions
is even more complex will become clear from the heat-capacity
measurements, which will be discussed later.

\begin{figure}[b]
\includegraphics[clip,width=65mm]{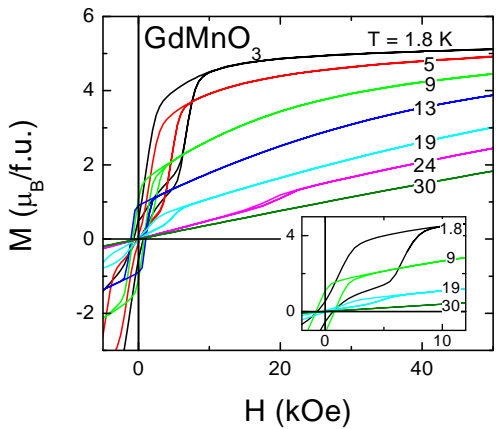}
\caption{Magnetization along $c$ as function of the external
magnetic field for a series of temperatures between 1.8~K and
30~K. The inset shows the hysteresis loops at low fields on an
enlarged scale.
\label{fig7}}
\end{figure}

The complex magnetization of GdMnO$_3$ as a function of field is
given in Fig.\ \ref{fig7}. An apparently paramagnetic
magnetization behavior at 30~K (we will see later that at 30~K
GdMnO$_3$  has to be characterized as an antiferromagnet with the
hysteresis shifted to higher fields), is followed at 24~K by a
$M(H)$ behavior, with the typical signature of a metamagnetic
phase transition at 20~kOe. On decreasing temperatures this
metamagnetic transition shifts towards lower fields and bears the
characteristics of a weak ferromagnet at 9~K. At the lowest
temperatures, below the ordering temperature of the Gd sublattice,
a complex hysteresis loop develops with a remnant magnetization
which however remains small and is well below $1~\mu_B$ even at
1.8~K. This complex hysteresis and the small remnant magnetization
signal show that at low fields ($H < 5$~kOe) and low temperatures
($T < 6$~K) both sublattices are canted and the ferromagnetic
components of the two sublattices are antiparallel to each other.
As the magnetization always increases with increasing field to
values well above the value of the ordered moment of Mn$^{3+}$, we
believe that the FM component of the Gd spins points in the
direction of the external field. At 50 kOe the magnetization of
GdMnO$_3$ reaches values slightly above $5~\mu_B$, hence the
manganese moments remain canted reducing the ordered moment of Gd
from $7~\mu_B$ to almost $5~\mu_B$. No indications for a further
spin reorientation (such as e.g.\ a spin-flop transition of the Mn
spins) and no increase of the magnetization above $M=6~\mu_B/$f.u.
can be detected in fields up to 140~kOe at $T=5$~K (not shown).

\begin{figure}[tb]
\includegraphics[clip,width=65mm]{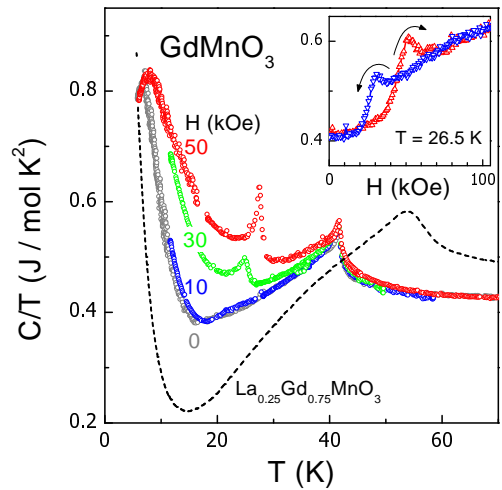}
\caption{Heat capacity of GdMnO$_3$ plotted as $C/T$ as a function
of temperature, measured at different external magnetic fields up
to 50~kOe. The dashed line represents data for
La$_{0.25}$Gd$_{0.75}$MnO$_3$ in zero external field. The inset
shows the field dependence of the heat capacity of GdMnO$_3$ at $T
= 26.5$~K.
\label{fig8}}
\end{figure}

Fig.\ \ref{fig8} presents the results of the heat-capacity
measurements in GdMnO$_3$ as function of temperature and field.
The heat capacity, plotted as $C/T$, is shown for temperatures $T
< 70$~K, in external magnetic fields up to 50~kOe. A well defined
heat-capacity anomaly, almost independent of the external field,
shows up close to 40~K. It has to be noted that at this
temperature no anomaly can be found in magnetic susceptibility
measurements. From this it becomes clear that this anomaly does
not result from the transition into the canted A-type AFM phase
but has to be interpreted as onset of an incommensurate AFM spin
structure\cite{kimura03}, which gives no macroscopic FM moment. A
further broad peak with a cusp-like anomaly appears at 7~K. On
increasing magnetic fields this low-temperature anomaly is shifted
slightly to higher temperatures and is smeared out gaining a much
higher weight of entropy between 7 and 40~K on increasing fields.
This Schottky-type contribution obviously results from the
alignment of the Gd spins within the external field amplified by
the effective field of the canted Mn-spin structure setting in at
$T_{\rm CA}\approx 20$~K, as known from the susceptibility
measurements. In fields of 30~kOe and 50~kOe a further anomaly
shows up close to 25~K, which is strongly dependent on the
external field. It probably corresponds to the metamagnetic
transition into a canted antiferromagnetic state visible in $M(H)$
in an external field (e.g.\ at 20~kOe and a temperature of 24~K,
as displayed in Fig.~\ref{fig7}). Presumably it merges with the Gd
anomaly at lower temperatures at lower fields (on the other hand
we cannot exclude a strong reduction of this anomaly with
decreasing magnetic field), corresponding to the observation that
the hysteresis loops are shifted to low magnetic fields at low
temperatures (Fig.~\ref{fig7}). For comparison the dashed line in
Fig.~\ref{fig8} exhibits the $C/T$ data for the composition
La$_{0.25}$Gd$_{0.75}$MnO$_3$. Here the above described complex
findings are different. At $T_{\rm N}=T_{\rm CA} \approx 55$~K a
cusp-like anomaly can be found which corresponds to the transition
into the canted AFM state (see Fig.~\ref{fig3}). A shoulder in the
displayed temperature dependence of $C/T$ below $T_{\rm CA}$ and
the strong increase towards low temperatures denote the
polarization of the Gd spins within the effective field of the Mn
sublattice. But no evidence for a further magnetic transition of
the Mn sublattice into an incommensurate phase was found.

\begin{figure}[b]
\includegraphics[clip,width=65mm]{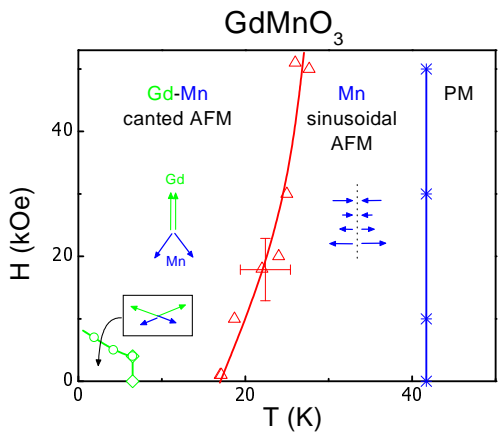}
\caption{$(H,T)$-phase diagram of GdMnO$_3$. The spin structures
of the two magnetic sublattices are indicated. \label{fig9}}
\end{figure}

From the ac- and dc-susceptibility measurements, the
magnetization, and the heat capacity we can construct a detailed
$(H,T)$-phase diagram which is shown in Fig.\ \ref{fig9}. The
paramagnetic regime of both sublattices above 40~K, is probably
followed by an incommensurate sinusoidal (or spiral) structure of
the manganese moments (in analogy to results for
TbMnO$_{3}$\cite{8} and o-HoMnO$_{3}$\cite{8a}) and an almost
paramagnetic gadolinium sublattice. Macroscopically any FM
component of the sinusoidal Mn order is averaged out. This means
that there is also no macroscopic net polarization of the Gd sites
induced by the sinusoidal structure of the manganese spins.
Towards lower temperatures this phase is followed by a canted
structure of the manganese moments exhibiting a small FM
component, due to which the Gd spins are polarized
antiferromagnetically. The FM contribution of the Gd spins is
larger than the FM contribution of the canted AFM state of the Mn
spins, directing the Gd moments parallel to the external field.
Only for sufficiently low temperatures and external magnetic
fields the Gd sublattice itself can setup an additional
antiferromagnetic (Gd-Gd) ordering, and we find a canted structure
for both Gd and Mn sublattices, still antiferromagnetically
coupled. For higher external magnetic fields this state is
suppressed.

\section{Conclusion}

We have presented a detailed study on single crystals of
La$_{1-x}$Gd$_x$MnO$_3$ covering the complete  concentration range
including pure GdMnO$_3$. Structural, magnetic, electrical, and
thermodynamic properties have been reported. The concentration
dependence of the lattice constants signals the increase of
orthorhombicity via  a stronger tilting and buckling. This results
from the decrease of the tolerance factor via the substitution of
the larger La by the smaller Gd ions. The overlaying JT distortion
and corresponding orbital order does not seem to be changed
considerably. At low temperatures the electronic charge transport
is dominated by hopping processes indicative for a disordered
semiconductor with a finite density of localized states. At
elevated temperatures all samples reveal a purely thermally
activated charge transport. The temperature of the JT transition
in GdMnO$_3$ could be estimated in the present work to be higher
than 1200~K. From the susceptibility and magnetization
measurements of the doped compounds we provide clear experimental
evidence for a ferromagnetic contribution of the magnetic order of
the manganese moments and for magnetic compensation due to the
increasing influence of the Gd moments which are
antiferromagnetically aligned in the internal field of the Mn
moments. The experimental observation of magnetic compensation and
spin reversal clearly indicates the importance of spin canting of
the manganese moments. Since the doping is isovalent,
double-exchange interactions cannot play any role and spin canting
probably results from the Dzyaloshinsky-Moriya
interaction.\cite{solovyev} This canted A-type AFM spin state
persists for low temperatures within the whole concentration
range.

Finally, for pure GdMnO$_3$ we observe a magnetic phase transition
at 40~K. We argue that below 40~K the manganese moments develop a
sinusoidal or another complex spin order. No polarization occurs
at the Gd site and the Gd spins behave paramagnetically. Below
16-20~K spin canting of the manganese moments occurs, polarizing
the Gd 4$f$ spins, and for further decreasing temperatures a weak
ferromagnetic moment evolves. Finally, at 6.5~K due to the
interaction of the 4$f$ spins, long-range order of the Gd moments
evolves, yielding a canting of the Gd spins with a FM component
antiparallel to the FM moment of the canted manganese spins. In
finite external fields this state is suppressed:  the canting
angle of the Gd-spins is reduced yielding the full Gd moment in
direction of the external field (reduced by the FM moment of the
canted Mn-lattice).

\acknowledgments

This work was partly supported by the Bundesministerium f\"ur
Bildung und Forschung (BMBF) via Grant No. VDI/EKM 13N6917-A, by
the Deutsche Forschungsgemeinschaft via  Sonderforschungsbereich
SFB 484 (Augsburg), and by INTAS and Russian Foundation for Basic
Researches N 03-02-16759.


\bibliographystyle{prsty}

\begin{references}
\bibitem{y-parameter} Even though the x-ray powder diffraction measurements
cannot reveal reliable oxygen positions, the Mn-O-Mn bond angles
could be estimated assuming a linear relation to the $y$-parameter
of the Gd sites (tilting distortion along $b$). The absolute
values for the bond angles (within the $ab$-plane) were scaled on
literature results\cite{kimura03} for the pure compounds LaMnO$_3$
and GdMnO$_3$.
\end{references}

\end{document}